\documentclass[twocolumn,spanish,aps,ams,showpacs,superscriptaddress]{revtex4}

\usepackage{graphicx,color}
\usepackage{latexsym}
\usepackage{amsmath,amssymb}        % amssymb includes amsfonts
\usepackage[draft=false]{hyperref}

\begin{document}

% -----> TITLE 

\title{Evolution of a mass-less test scalar field on Boson Stars space-times}

% ----->   AUTHORS   <-----

\author{F. D. Lora-Clavijo}
\affiliation{Instituto de F\'{\i}sica y Matem\'{a}ticas, Universidad
              Michoacana de San Nicol\'as de Hidalgo. Edificio C-3, Cd.
              Universitaria, 58040 Morelia, Michoac\'{a}n,
              M\'{e}xico.}
              
\author{A. Cruz-Osorio}
\affiliation{Instituto de F\'{\i}sica y Matem\'{a}ticas, Universidad
              Michoacana de San Nicol\'as de Hidalgo. Edificio C-3, Cd.
              Universitaria, 58040 Morelia, Michoac\'{a}n,
              M\'{e}xico.}

\author{F. S. Guzm\'an}
\affiliation{Instituto de F\'{\i}sica y Matem\'{a}ticas, Universidad
              Michoacana de San Nicol\'as de Hidalgo. Edificio C-3, Cd.
              Universitaria, 58040 Morelia, Michoac\'{a}n,
              M\'{e}xico.}
              
% --->   DATE

\date{\today}

% -----> ABSTRACT

\begin{abstract}
We numerically solve the massless test scalar field equation on the space-time background of  boson stars and black holes. In order to do so, we use a numerical domain that contains future null infinity. We achieve this construction using a scri-fixing conformal compactification technique based on hyperboloidal constant mean curvature foliations of the space-time and solve the conformally invariant wave equation. We present two results: the scalar field shows oscillations of the quasinormal mode type found for black holes only for boson star configurations that are compact; and no signs of tail decay are found in the parameter space we explored. Even though our results do not correspond to the master equation of perturbations of boson star solutions, they indicate that  the parameter space of boson stars as black hole mimickers is restricted to compact configurations.
\end{abstract}

% ----->   PACS


% ----->   MAKETITLE   <-----

\maketitle

% ----->     INTRODUCTION     <-----

\section{Introduction}

Observations related to high energy events, indicate the existence of black hole candidates (BHCs). It is usually assumed that black hole solutions are the models of BHCs. Nevertheless there are various nonvacuum solutions in general relativity that are being studied as alternatives to black holes, sometimes called black hole mimickers. Among such alternative solutions are wormholes \cite{Harko2009,Lemos2008}, gravastars \cite{Visser2004,Rezzolla2007,Lemos2008,Harko2009c}, brane world solutions \cite{Harko2008} and boson stars \cite{diego,diego-acc,Guzman2006,GuzmanBecerril2009}. Predictions on phenomena related to black hole mimickers are important, either because they may rule out the mimicker or at least restrict the parameter space of mimicker configurations.

One of the most important properties of mimickers is the stability. For wormhole solutions instability  has been shown for basic wormhole solutions \cite{Hayward,GGS1,GGS2,GGS3}, destroying previous hope on the possibility that these are stable as shown in \cite{Armendariz}. The stability of gravastars has been explored and it has been found that there are stability regions \cite{Visser2004,Cardoso,Carter2005,Rezzolla2008}. In favor of boson stars  the stability of the solutions has been exhaustively studied, using perturbative methods \cite{Gleiser,scott}  and using full nonlinear numerical relativity, both in spherical symmetry  \cite{SeidelSuen1990,Balakrishna1998} and full three dimensions \cite{Guzman2004}.

In the particular case of boson stars they have been studied as sources of gravitational radiation, both as binary systems \cite{Lehner2007} and as perturbed boson stars \cite{Balakrishna2006} in the full nonlinear regime. The study of gravitational wave signatures has also been pursued in order to differentiate between a black hole and a gravastar \cite{Rezzolla2007}. Instead, in the case of wormholes for example, simple solutions (supported by a phantom scalar field) would not stand the perturbative analysis and there is no hope for a study of a binary system,  because as shown in \cite{GGS2} the lifetime of these solutions is rather short and they should either collapse and form black holes or explode before they could merge (although in \cite{GGS3} it was shown that  charged wormholes can have a longer lifetime if the charge parameter is adequately chosen). At the end of the day, if black hole mimickers are to be compared from all the angles with black holes, mimickers should also be expected to do what black holes can do: they collide and generate gravitational radiation with a given fingerprint.

In this paper we explore a simple but potentially fruitful problem: the evolution of a test massless scalar field on the space-time of boson stars. It is well known that, in the case of black holes, the solution of such an equation is related to the quasinormal modes of axial perturbations and electromagnetic test fields \cite{nollert}. And even if this equation is not the perturbation equation of boson stars (it would involve the perturbation of matter too) the behavior of the massless scalar field would provide indications on which boson stars look more like black holes, among the big set of boson stars that are mimickers when the power emitted by accretion disks are compared to those due to black holes \cite{Guzman2006,GuzmanBecerril2009}. We calculate the numerical solution for various boson star space-times and compare to the solution for the space-time corresponding to a Schwarzschild black hole. We track the numerical solution at the level of mode ringing and tail decay.

One of the important ingredients of this paper is that we foliate our space-times with hyperboloidal slices,  for which we develop the procedure to foliate a static spherically symmetric space-time with constant mean curvature slices. Following \cite{anil1,anil2}, we compactify the boundary at ${\cal J}^{+}$ (future null infinity) which allows one to measure the amplitude of the scalar field at almost arbitrary distances from the source. The compactification of future null infinity implies the physical metric to be singular at infinity; this problem is fixed using a rescaling through a conformal transformation \cite{anil1}. Therefore, in order to achieve correct physical results, we solve the conformally invariant wave equation. We proceed in this manner because we are interested in measuring the scalar field far from the source and at infinity; in this case it is natural to consider future null infinity as the boundary because it is the asymptotic boundary of a scalar field propagating at the speed of light.

The paper is organized as follows. In Sec. II we describe the construction of boson star space-times. In Sec. III we develop the process of conformal compactification for a spherically symmetric space-time, that we apply to the Schwarzschild and boson star space-times. In Sec. IV we present the solution of a massless  scalar field on these space-times and compare the behavior by solving the conformally invariant wave equation. Finally in Sec. V we present some conclusions.

% ----------     Boson Stars SECTION II

\section{Boson stars} 
\label{sec:BSs}

Boson stars (BSs) arise from the Lagrangian density of a complex scalar field minimally coupled to
gravity, that is,

\begin{equation}
{\cal L} = -\frac{R}{\kappa_0} + g^{\mu \nu}\partial_{\mu} \Phi^{*}
\partial_{\nu}\Phi + V(|\Phi|^2),
\label{eq:lagrangian}
\end{equation}

\noindent where $\kappa_0 = 16 \pi G$ in units where $c =1$, $\Phi$ is the
scalar field, the star
stands for complex conjugate, and $V$ the potential of
self-interaction of the field \cite{jetzer,topical-review}. Notice that this Lagrangian density
is invariant under the global $U(1)$ group, and the associated
conserved charge is the quantity called the number of particles (defined below). When
the action is varied with respect to the metric,
Einstein's equations arise $G_{\mu\nu} = \kappa_0 T_{\mu\nu}$, where the
resulting stress-energy tensor reads

\begin{equation}
T_{\mu \nu} = \frac{1}{2}[\partial_{\mu} \Phi^{*} \partial_{\nu}\Phi +
\partial_{\mu} \Phi \partial_{\nu}\Phi^{*}] -\frac{1}{2}g_{\mu \nu}
[\Phi^{*,\alpha} \Phi_{,\alpha} + V(|\Phi|^2)].
\label{eq:set}
\end{equation}

\noindent Boson stars are related to the potential
$V=m^2 |\Phi|^2 + \frac{\lambda}{2}|\Phi|^4$. The quantity $m$ is
understood as the mass of a boson and $\lambda$ is the coefficient of a
two-body self-interaction mean field approximation. The Bianchi identity
reduces to the Klein-Gordon equation

\begin{equation}
\left(
\Box - \frac{dV}{d|\Phi|^2}
\right)\Phi = 0,
\end{equation}

\noindent where $\Box
\Phi=\frac{1}{\sqrt{-g}}\partial_{\mu}[\sqrt{-g}
g^{\mu\nu}\partial_{\nu}\Phi]$.

Boson stars are spherically symmetric solutions to the above set of equations under a
particular condition: the scalar field has harmonic time
dependence $\Phi(r,t) = \phi_0(r) e^{-i \omega t}$, where $r$ is the
radial spherical coordinate. This condition implies that the stress-energy 
tensor in (\ref{eq:set}) is time-independent, which implies
through Einstein's equations that the geometry is also time-independent.
That is, there is a time-dependent scalar field oscillating upon a
time-independent geometry whose source is the scalar field itself.
It is possible to construct solutions for boson stars assuming that
the metric can be written in normal coordinates as

\begin{equation}
ds^2=-\alpha(r)^2dt^2 + a(r)^2dr^2 + r^2 d\Omega^2.
\label{eq:spherical_metric}
\end{equation}

\noindent For these coordinates the Einstein-Klein-Gordon system of
equations reads:

\begin{eqnarray}
\frac{\partial_r a}{a} &=& \frac{1-a^2}{2r} +\nonumber\\
        &&\frac{1}{4}\kappa_0 r
        \left[\omega^2 \phi_{0}^{2}\frac{a^2}{\alpha^2}
        +(\partial_r \phi_0)^{2} +
        a^2 \phi_{0}^{2} (m^2
        + \frac{1}{2}\lambda \phi_{0}^{2})
        \right],\nonumber\\%\label{sphericalekga-sc}\\
\frac{\partial_r \alpha}{\alpha} &=&
        \frac{a^2-1}{r} +
        \frac{\partial_r a}{a} -
        \frac{1}{2}\kappa_0 r a^2\phi_{0}^{2}(m^2
        +\frac{1}{2}\lambda\phi_{0}^{2}),\nonumber\\%\label{sphericalekgb-sc}\\
\partial_{rr}\phi_0  &+& \partial_r \phi_0  \left( \frac{2}{r} +
        \frac{\partial_r \alpha}{\alpha} - \frac{\partial_r a}{a}\right)
        + \omega^2 \phi_0 \frac{a^2}{\alpha^2} \nonumber\\
        &-& a^2 (m^2 + \lambda \phi_{0}^{2}) \phi_0
        =0.\label{sphericalekgc-sc}
\end{eqnarray}

\noindent The system
(\ref{sphericalekgc-sc}) is a set of
coupled ordinary differential equations to be solved under the
conditions of spatial flatness at the origin $a(0)=1$, $\phi_0(0)$
finite and $\partial_r \phi_0(0)=0$ in
order to guarantee regularity and spatial flatness at the origin, and
%$\phi_0(\infty)=\phi_0 \prime(\infty)=0$ in order to ensure asymptotic
$\phi_0(\infty)=0$ in order to ensure asymptotic
flatness at infinity as described in
\cite{ruffini,SeidelSuen1990, Balakrishna1998,Gleiser}; these
conditions reduce the system
(\ref{sphericalekgc-sc}) to an
eigenvalue problem for $\omega$, that is, for every central value of
$\phi_0$ there is a unique $\omega$ with which the boundary conditions
are satisfied. In practice the system of equations is not integrated up to infinity, instead the solution is calculated up to a large value of $r$ where the boundary conditions are satisfied with a given tolerance.
 Because the system
(\ref{sphericalekgc-sc})
contains several constants, it is convenient to rescale the variables in such
a way that they do not appear. In order to do so,
the following transformation is convenient:
$\hat{\phi}_0 = \sqrt{\frac{\kappa_0}{2}} \phi_0$,
$\hat{r} = mr$,
$\hat{t} = \omega t$,
$\hat{\alpha} = \frac{m}{\omega}\alpha$ and
$\Lambda = \frac{2\lambda}{\kappa_0 m^2}$ \cite{Balakrishna1998}. The result is that the physical
constants vanish from the equations and the radial coordinate has
units of $m$ and the time has units of $\omega$. In fact the
mass of the boson becomes the parameter that fixes the scale of the system.
After substituting this transformation and removing the tildes from
everywhere, the resulting Einstein-Klein-Gordon system of equations reads:

\begin{eqnarray}
\frac{\partial_r a}{a} &=& \frac{1-a^2}{2r} +\nonumber\\
        &&\frac{1}{2} r
        \left[\phi_{0}^{2}\frac{a^2}{\alpha^2}
        +(\partial_r \phi_0)^{2} +
        a^2 (\phi_{0}^{2}
        + \frac{1}{2}\Lambda \phi_{0}^{4})
        \right],\nonumber\\%\label{sphericalekga-sc}\\
\frac{\partial_r \alpha}{\alpha} &=&
        \frac{a^2-1}{r} +
        \frac{\partial_r a}{a} -
        r a^2\phi_{0}^{2}(1
        +\frac{1}{2}\Lambda\phi_{0}^{2}),\nonumber\\
\partial_{rr}\phi_0  &+& \partial_r \phi_0
        \left(
                \frac{2}{r} +
                \frac{\partial_r \alpha}{\alpha} - \frac{\partial_r a}{a}
        \right)
        + \phi_0 \frac{a^2}{\alpha^2} \nonumber\\
        &-& a^2 (1 + \Lambda \phi_{0}^{2}) \phi_0
        =0.
\label{sphericalekgc-sc-rescaled}
\end{eqnarray}

\noindent Notice that the parameter $\omega$ now turns into
the central value of the lapse $\alpha$ due to the rescaling. This is the
system that is being solved in practice using finite differences with an ordinary
integrator (adaptive step-size fourth order Runge-Kutta algorithm in the present case) and a shooting routine
that bisects the value of $\omega$.

% FIGURE

\begin{figure}[htp]
\includegraphics[width=8cm]{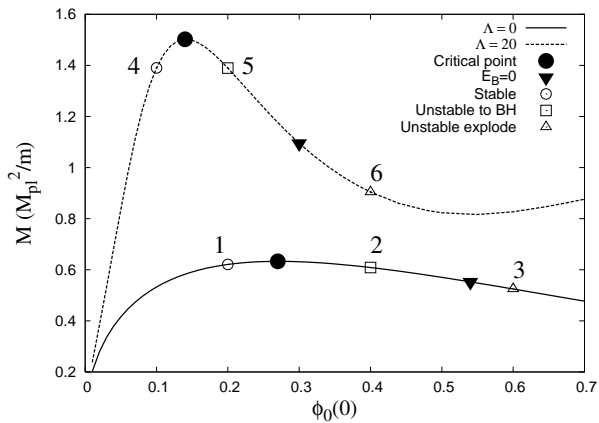}
\caption{\label{fig:equilibrium} Sequences of equilibrium
configurations for two values of $\Lambda$ are shown as a function
of the central value of the scalar field $\phi_0(0)$; each point in the curves
corresponds to a solution of the eigenvalue problem and represents a boson star
configuration. The filled circles indicate the critical solution that
separates the stable from the unstable branch. Those
configurations to the left of the maxima represent stable
configurations. The inverted triangles
indicate the point at which the binding energy is zero. Those
configurations between the filled circles and the inverted triangles (along each
sequence) collapse into black holes as a response to a perturbation.
Configurations to the right of the inverted triangles disperse away.}
\end{figure}

The solutions of (\ref{sphericalekgc-sc-rescaled}) define sequences of
equilibrium configurations like those shown in Fig.
\ref{fig:equilibrium}. Each point in the curves corresponds to a
boson star solution. In each of the curves two
important points for each value of $\Lambda$ are marked: (i) the critical
point, marked with a filled circle, indicating the threshold between the
stable and unstable branches of each sequence (that is, configurations
to the left of this point are stable and those to the right are unstable)
as found through the analysis of perturbations \cite{Gleiser} and full nonlinear evolution of
the equilibrium solutions
\cite{SeidelSuen1990,Balakrishna1998,Guzman2004};
and (ii) the point at which the binding energy $E_B = M-Nm = 0$ marked
with an inverted filled triangle (see \cite{Gleiser} for this convention of the binding
energy), where $N=\int
j^0 d^3 x = \int
\frac{i}{2}\sqrt{-g}g^{\mu\nu}[\Phi^{*}\partial_{\nu}\Phi - \Phi
\partial_{\nu}\Phi^{*}]d^3 x$ is the number of particles; that is,
the conserved quantity due to the invariance under the global $U(1)$
group of the Lagrangian density (\ref{eq:lagrangian}). $M =
(1-1/a^2)r/2$ evaluated at the outermost point of the numerical domain
is our best guess of the Arnowitt-Deser-Misner (ADM) mass \cite{Balakrishna1998}; the configurations between the instability
threshold and the zero binding energy point (like configurations 2 and 5) have negative binding energy
($E_B<0$) and collapse into black holes whereas those to the right (like 3 and 6)
have positive binding energy and disperse away. Those configurations to the left of the threshold of instability like 1 and 4, that is,
stable configurations, possess a negative binding energy.
Boson stars as black hole mimickers are located in the stable branch \cite{Guzman2006,GuzmanBecerril2009}, and that is why in this paper we only consider stable configurations.

% --------     Conformal compactification SECTION III

\section{Scri-fixing conformal compactification}

We want to solve for a test scalar field on a fixed background geometry corresponding to a boson star and compare it with the one calculated on a Schwarzschild background. For this we choose hyperboloidal  slices that show important advantages.
At present time, foliations with boundaries at future null infinity are being used for various applications, like the solution of perturbation equations \cite{whs,dario,anil2009} and the study of tails \cite{tiglio}. 
On the other hand, hyperboloidal foliations are useful because they reach future null infinity and for asymptotically flat space-times, the gravitational radiation and scalar field global properties are well defined at such a boundary \cite{bondi,sachs}. On the other hand, it is possible to compactify the space-time in order to work on a numerical domain that contains ${\cal J}^{+}$, for which it is necessary to compactify the spatial coordinates in an appropriate way such that the space-time is regular there, at future null infinity. %Some foliations with slices that reach future null infinity can also  behave like the conventional 3+1 foliations near the source and approach the future null infinity far from the source \cite{stewart,Frie,Frau}. These foliations are called hyperboloidal \cite{Frie}. 
%Other important consequence of having future null infinity in a numerical domain is that it is possible to compute the correct Bondi mass of a system \cite{bondi}, which is a specially important quantity for systems containing scalar fields. For these reasons, having foliations that reach future null infinity is preferred in this work. 

Usually, when a spatial coordinate is being compactified, the metric is singular at the new spatial infinity, and the same happens when hyperboloidal slices are compactified. Among others \cite{rinnemoncrief,rinne}, a  known way of fixing this problem is the definition of a conformal metric that uses an {\it ad hoc} conformal factor that regularizes the singular terms resulting from the compactification \cite{anil1,rinne}. The idea behind  scri-fixing conformal compactification for spherically symmetric space-times is the following: (i) use hyperboloidal foliations whose slices reach ${\cal J^{+}}$, (ii) compactify the radial coordinate and (iii) regularize the metric with an appropriate conformal factor. In order to proceed, a new time coordinate is introduced as $t = \tilde{t} - h(\tilde{r})$, where ($\tilde{t},\tilde{r}$) are the original time and space coordinates and $h(\tilde{r})$ is a function called the height function \cite{anil1,rinne,Murc}. This last transformation has the 
advantage that it keeps the time direction invariant. That is, regardless of the choice of the $h(\tilde{r})$,  the timelike Killing vector has the same form in both coordinate systems.
A compactifying coordinate is introduced in the form  $\tilde{r} =  \frac{r}{\Omega}$, where $\Omega=\Omega(r)$ is a function that compactifies $r$ and at the same time is a non-negative conformal factor that vanishes at  ${\cal J^{+}}$ \cite{anil1}; in our case we choose $\Omega=1-r$ for the black holes and $\Omega = (1-r^2)/2$ for BSs. Then $\cal J^{+}$ is  located at $r=1$, where the space-time can be rescaled $g=\Omega^{2}\tilde{g}$ with the conformal factor so that the conformal metric $g$ is regular everywhere. 

On the other hand, for the construction of $h(\tilde{r})$, constant mean curvature slices are used \cite{Murc}. The mean extrinsic curvature $\tilde{k}$  of the initial slice $t=0$ is given by \cite{Murc}

\begin{equation}
\tilde{k} = \nabla_{\mu}n^{\mu} = \frac{1}{\sqrt{-g}}\partial_{\mu}(\sqrt{-g}n^{\mu}), \label{eq:curv}
\end{equation}

\noindent where $n^{\mu}$ is a timelike unit vector normal to the spatial hypersurface and positive $\tilde{k}$ means that the slices reach future null infinity. In this paper, we adopt the condition that the mean extrinsic curvature $\tilde{k}$ is constant. Such condition allows the integration of the equation above.

We now develop the construction of the conformal compactification for a spherically symmetric space-time described by the type of metric we start up in the cases studied here (Schwarzschild and the boson star):

\begin{equation}
d\tilde{s}^2 = -\alpha^2(\tilde{r}) d\tilde{t}^2 + a^2(\tilde{r}) d\tilde{r}^2 +
	\tilde{r}^2 (d\theta^2 + \sin^2 \theta d\varphi^2),
\label{eq:Boson-metric}
\end{equation}

\noindent where $a$ and $\alpha$  are assumed to be known metric functions, and we proceed to construct a hyperboloidal foliation and a scri-fixing compactification.

By introducing the change of coordinate $t=\tilde{t} - h(\tilde{r})$  the line element takes the form

\begin{eqnarray}\label{eq:slicing}
\nonumber && d\tilde{s}^{2}=-\alpha^2(\tilde{r})dt^{2} - 2h'(\tilde{r})\alpha^2(\tilde{r})dtd\tilde{r} \nonumber\\ 
&& + \left[a^2(\tilde{r})  - \alpha^2(\tilde{r}) (h'(\tilde{r}))^2 \right]d\tilde{r}^{2} +  \tilde{r}^{2}(d\theta^{2} + \sin^2 \theta d\phi^{2}), 
\end{eqnarray}

\noindent where $h^{\prime}=\frac{dh}{d\tilde{r}}$. Comparison of this metric with the 3+1 metric 
$d\bar{s}^2 = (-\bar{\alpha}^2 + \bar{\gamma}^2 \bar{\beta}^2)dt^2 + 2 \bar{\beta}\bar{\gamma}^2 d \tilde{r}dt + \bar{\gamma}^2 d\tilde{r}^2 + \tilde{r}^2 (d\theta^2 + \sin^2 \theta d\varphi^2)$ allows one to read off the gauge and metric functions

\begin{eqnarray}
\bar{\alpha}(\tilde{r}) &=& \frac{\alpha(\tilde{r}) a(\tilde{r})}{\bar{\gamma}(\tilde{r})},\nonumber\\
\bar{\beta}(\tilde{r}) &=& -\frac{h'(\tilde{r})\alpha^2(\tilde{r})}{\bar{\gamma}^{2}(\tilde{r})},\nonumber\\
\bar{\gamma}^{2}(\tilde{r}) &=& a^2(\tilde{r}) - \alpha^2 (\tilde{r})(h'(\tilde{r}))^2.
\label{eq:gauge_sch}
\end{eqnarray}

In terms of these functions, the unit normal vector to the spatial hypersurfaces pointing to the future is given by

\begin{equation}
n^{\mu} = \left[ \frac{\bar{\gamma}(\tilde{r})}{\alpha(\tilde{r})a(\tilde{r})}, \frac{h'(\tilde{r})\alpha(\tilde{r})}{\bar{\gamma}(\tilde{r})a(\tilde{r})}, 0 ,0 \right].
\end{equation}

Given a spacelike slice, we can drag it along the timelike Killing vector. This  will give a slicing where the time translation is along the Killing vector. Now, the mean curvature at the initial time slice, given by (\ref{eq:curv}), takes the form

\begin{equation}
\tilde{k} = \frac{1}{\tilde{r}^{2}\alpha(\tilde{r})a(\tilde{r})}\partial_{\tilde{r}}\left[\frac{\tilde{r}^{2}h'(\tilde{r})\alpha^2 (\tilde{r})}{\bar{\gamma}(\tilde{r})}\right],
\end{equation}

\noindent and can be integrated for constant $\tilde{k}$:

\begin{equation}
\tilde{k}\int \tilde{r}^2 \alpha(\tilde{r})a(\tilde{r})d\tilde{r} - C = \frac{\tilde{r}^{2}h'(\tilde{r})\alpha^2 (\tilde{r})}{\bar{\gamma}(\tilde{r})},
\end{equation}

\noindent where now $h'(\tilde{r})$ is

\begin{equation} \label{eq:h'}
h'(\tilde{r}) = \frac{[\tilde{k}I(\tilde{r}) - C]a(\tilde{r})}{\alpha(\tilde{r})\sqrt{[\tilde{k}I(\tilde{r}) - C]^2 + \alpha^2(\tilde{r})\tilde{r}^4}}
\end{equation}

\noindent and 

\begin{equation}
I(\tilde{r}) = \int \tilde{r}^2 \alpha(\tilde{r})a(\tilde{r})d\tilde{r}.
\end{equation} 

\noindent In this case, it is not easy to find $h$ in a closed form, so in order to have a description of the slices in general one has to integrate this function numerically. 

On the other hand, in order to perform the scri-fixing compactification,  we define the compact coordinate $r$ by $\tilde{r}=\frac{r}{\Omega}$ and we rescale the original metric by the conformal factor $\Omega$. The space-time using scri-fixing conformal compactification is finally given by the line element 

\begin{eqnarray}
\nonumber && ds^{2}=-\alpha^2(r)\Omega^2(r) dt^{2} - 2h'(r)\alpha^2(r)(\Omega - r\Omega')dtdr \nonumber\\ 
&& + \left[a^2(r)  - \alpha^2(r) (h'(r))^2 \right](\Omega-r\Omega')^2\frac{dr^{2}}{\Omega^2(r)}  \nonumber \\ 
&& +  r^{2}(d\theta^{2} + \sin^2 \theta d\varphi^{2}), 
\label{eq:Sph_compactifiedgeneral}
\end{eqnarray} 
 
\noindent where the functions $\alpha(r)$, $a(r)$ and $h'(r)$  are the functions $\alpha(\tilde{r})$, $a(\tilde{r})$ and $h'(\tilde{r})$ evaluated in $\tilde{r}=\tilde r (r)$. The conformal factor $\Omega$ determines various properties of the resulting conformal metric.

The two cases we study here correspond to the Schwarzschild and the boson star space-times. In the first case, $\alpha^2 = \frac{1}{a^2}= (1-\frac{2M}{\tilde{r}})$, with which our expression (\ref{eq:h'}), using $\Omega=1-r$, is reduced to the expression for $h'(\tilde{r})$ obtained by Malec and Murchadha \cite{Murc}, and the final version of the conformally rescaled metric reads

\begin{eqnarray}
ds^{2}&=&-\left(1-\frac{2M\Omega}{r}\right)\Omega^{2}dt^{2}-\frac{2(\tilde{k}r^{3}/3 -C\Omega^{3})}{P(r)}dtdr \nonumber\\
&+& \frac{r^{4}}{P^{2}(r)} dr^{2}+ r^{2}(d\theta^{2} + \sin^2 \theta d\phi^{2}),
\label{eq:Sch_compactified}
\end{eqnarray}

\noindent where

\begin{equation}\label{eq:ppp}
P(r)=\Omega^{3}\tilde{P}(r)= \sqrt{\left(\frac{\tilde{k}r^{3}}{3}-C\Omega^{3}\right)^{2} + \left(1-\frac{2M\Omega}{r}\right)\Omega^{2}r^{4}}.
\end{equation}

\noindent The values of $\tilde{k}$ and the integration constant $C$ are restricted such that $P(r)$ is real. The values of these constants used in this paper are $\tilde{k} = 0.4$ and $C=2$ in all our calculations for the black hole.

In the Boson Star case: 
(i) we solve the eigenvalue problem (\ref{sphericalekgc-sc-rescaled}) for a given configuration with a given ADM mass; 
(ii) since such solution is only calculated up to a finite -but large- radius, we match the solution to the Schwarzschild solution;
(iii) then we foliate the resulting solution using hyperboloidal slices with constant mean curvature using (\ref{eq:h'});
and (iv)  finally we construct a conformal metric that contains a compactified spatial coordinate with the values $\tilde{k}=3$ and $C=0$, which we have found to be useful for space-times without horizons.

Since boson star space-times are regular everywhere and since we solve the conformally invariant wave equation where the Ricci scalar gets involved, it is useful to use an adequate conformal factor that allows such scalar to be regular at the origin. Following \cite{anilphi3} we use $\Omega=(1-r^2)/2$ which allows a finite value of the Ricci scalar at the origin for the Minkowski space-time and we found this to be useful in the BS case too.

%     ------ Solution of the conformal wave equation SECTION IV

\section{Solution of the wave equation}

Following our convention, the conformal metric is given by $ds^2 = \Omega^2 d\tilde{s}^2$, meaning that the physical metric is the one with the tildes. We decide to solve the wave equation using the conformal metric because of various reasons: (i) the slices are hyperboloidal and reach ${\cal J}^{+}$ at infinity; (ii) the spatial coordinate is compactified, and therefore the wave function reaches future null infinity at the boundary $r=1$; (iii) such boundary is null and there is no need to impose boundary conditions there.

In order to use these benefits it is necessary to solve the conformally invariant wave equation

\begin{equation}
\left( \Box - \frac{1}{6}R \right)\phi_T(t,r,\theta,\varphi) = 0,
\label{eq:conf_inv}
\end{equation}

\noindent where $R$ is the Ricci scalar of the conformal metric, $\Box=\nabla^{\mu}\nabla_{\mu}$ also corresponds to the conformal metric and the physical scalar field $\tilde{\Phi}$ is related to the conformal scalar field by $\tilde{\Phi} = \Omega \Phi$. In order to study nonradial modes, we write separate $\phi_T(t,r,\theta,\varphi)=\phi(t,r) Y_{lm}(\theta,\varphi)$ with $Y_{lm}$ the spherical harmonics and solve the resulting equations.  We solve numerically this equation considering a domain $r \in [r_{min},1]$. For the case of the black holes we choose $r_{min}$ such that it lies inside the event horizon and satisfies the need of $P(r)$ being real. For the case of the boson stars, since such solutions are real everywhere we choose $r_{min}=0$.

We solve (\ref{eq:conf_inv}) as an initial value problem using a first order variable formulation. In terms of the line element (\ref{eq:Sph_compactifiedgeneral}) we can read off again the gauge in terms of the ADM-like metric $d\hat{s}^2 = (-\hat{\alpha}^2 + \hat{\gamma}^2 \hat{\beta}^2)dt^2 + 2 \hat{\beta}\hat{\gamma}^2 drdt + \hat{\gamma}^2 dr^2 + r^2 (d\theta^2 + \sin^2 \theta d\varphi^2)$ and obtain the following gauge and metric functions

\begin{eqnarray}
\hat{\alpha}^{2}(r) &=& \alpha^2(r) \Omega^2 + \hat{\beta}^2(r) \hat{\gamma}^2(r),\nonumber\\
\hat{\beta}(r) &=& -\frac{\alpha^2(r) h'(r) (\Omega - r\Omega')}{\hat{\gamma}^2(r)},\nonumber\\
\hat{\gamma}^2(r) &=& \frac{(a^2(r) - \alpha^2(r) h'^2(r))(\Omega-r\Omega')^2}{\Omega^2},\nonumber\\
\label{eq:gauge_sch}
\end{eqnarray}

\noindent which are the final metric functions we use to solve the wave equation. We choose to solve (\ref{eq:conf_inv}) as a first-order system, for which we define first-order variables $\pi = \frac{\hat \gamma}{\hat \alpha}\partial_t \phi - \frac{\hat \gamma}{\hat \alpha} \hat \beta \partial_r \phi$ and $\psi = \partial_r \phi$. Then Eq. (\ref{eq:conf_inv}) is transformed into the following system of equations:

\begin{eqnarray}
\partial_t \psi&=&  \partial_r\left( \frac{\hat \alpha}{\hat \gamma}\pi + \hat \beta \psi \right),\nonumber\\
\partial_t \pi&=& \frac{1}{r^2} \partial_r \left(r^2 (\hat \beta \pi + \frac{\hat \alpha}{\hat \gamma}\psi ) \right) - \hat \alpha \hat \gamma \left( \frac{1}{6}R\phi + \frac{l(l+1)}{r^2}\phi \right),
\label{eq:1st_order}
\end{eqnarray}

\noindent where we have included the separation of the angular part of the scalar field in  (\ref{eq:conf_inv}).
Finally, we integrate the system of equations (\ref{eq:1st_order}) numerically using a finite differences approximation on a uniformly discretized domain; we use the method of lines with sixth-order accurate stencils along the spatial direction and a fourth-order Runge-Kutta time integrator.

We use initial data corresponding to an outgoing Gaussian profile for the scalar field:

\begin{eqnarray}
\phi(0,r) &=&  Ae^{-(r-r_{0})^2/\sigma^2}, \nonumber \\
\psi(0,r) &=& -2 \frac{(r-r_{0})}{\sigma^2}\phi(0,r),\nonumber\\
\pi (0,r) &=& -\psi(0,r) - \frac{\phi(0,r)}{r}\left( 1 - \frac{\hat \beta \hat \gamma}{\hat \alpha}\right). \label{eq:initial_data}
\end{eqnarray}

At the boundary $r=1$ we do not need to impose boundary conditions as mentioned above, however at $r=r_{min}$ we do the following: (i) in the case of the black holes, since this boundary is located inside the event horizon, we do not need to impose boundary conditions given that the cones point inward there and we only verify that no spurious signals propagate out of the horizon; and  (ii) in the case of boson stars, we have to deal with the origin $r=0$, where we first stagger the origin, and second we substitute the term $\frac{1}{r^2}\frac{\partial f}{\partial r}$ by $3\frac{\partial f}{\partial r^3}$ in (\ref{eq:1st_order}).

\subsection{Tests}

\begin{figure}[htp]
\includegraphics[width=8cm]{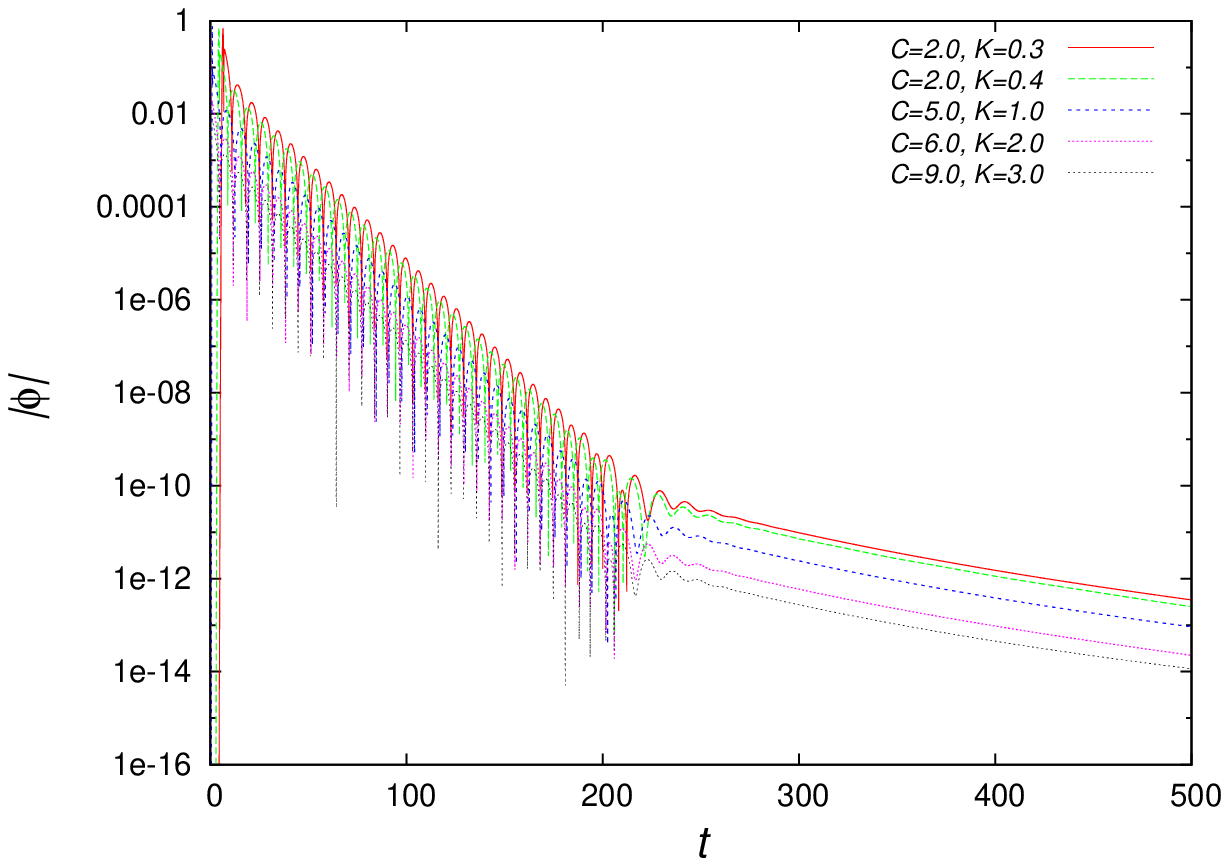}
\includegraphics[width=8cm]{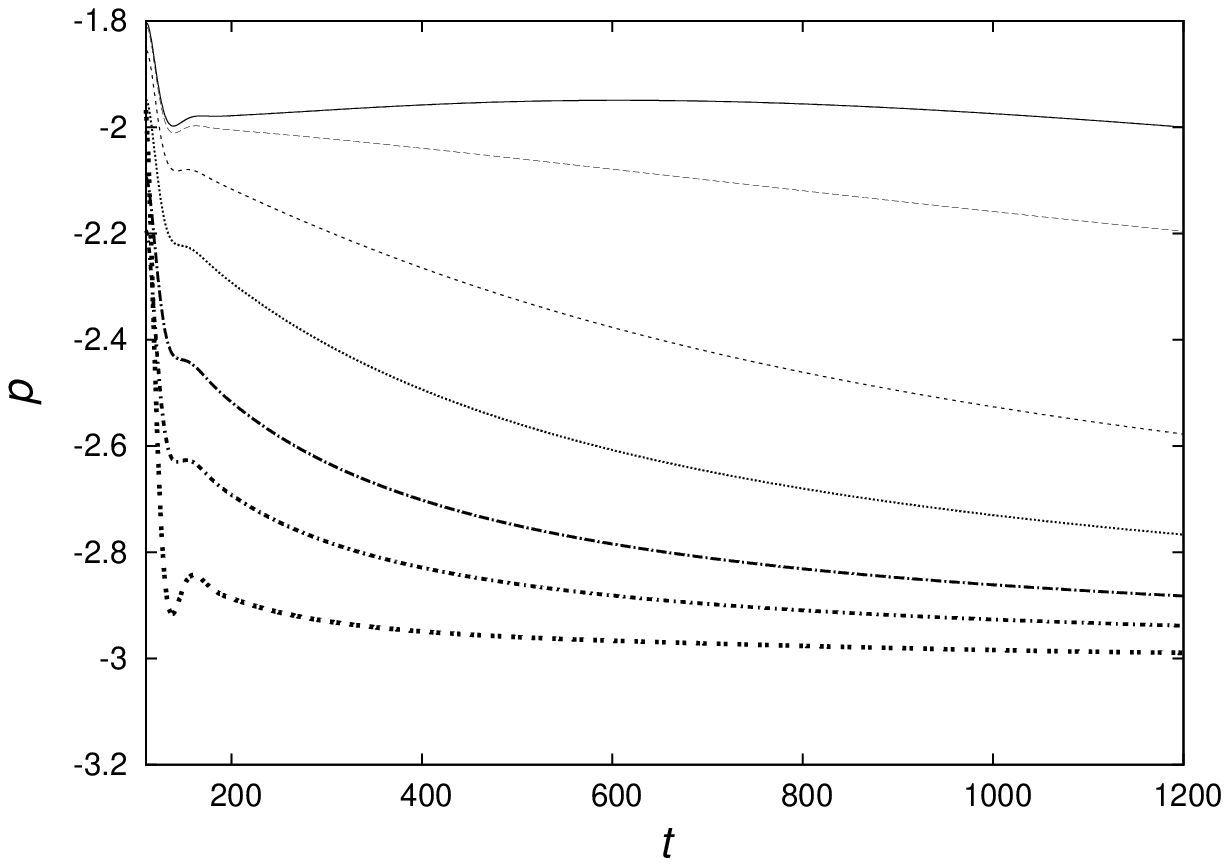}
\caption{\label{fig:test1} (Top) We show the absolute value of the scalar field amplitude measured at a fixed radius $r=1000M$, for various values of the foliation parameters for a black hole of mass $M=1/2$, and mode $l=2$. We found that the quasinormal modes have a frequency $\omega = 0.749\pm0.004$ which is consistent with previous studies than indicate $\omega =0.7473$ \cite{Leaver}, and the tail decay exponent are both independent of the foliation parameters. (Bottom) We show the tail decay exponent for $l=0$ to be of the form $\sim t^p$  measured also at various distances from the black hole's horizon. These results are consistent with the predictions of measurements by timelike observers ($p=-3$ \cite{p3}) and null observers at ${\cal J}^{+}$ ($p=-2$ \cite{p2}), and also with recent numerical calculations of the polynomial exponent \cite{anil2}. The curves represent the measurment by detectors located at: infinity, $1000M$, $250M$, $110M$, $50M$, $24M$ and $5M$, from top to bottom, respectively.} 
\end{figure}

We use the solution of the test massless scalar field on the Schwarzschild solution in order to have a first validation of our implementations. For this we choose the case of a Schwarzschild black hole solution with mass and show the amplitude of the scalar field  in Fig \ref{fig:test1}, where the quasinormal mode and the tail decay for $l=2$ are shown and validated. From such result we fit the frequency of the oscillation which is consistent with the results in \cite{Leaver}. 
%The tail decay rates measured following the quasinormal mode ringing phase $\phi \sim 1/t^p$, was found to be within $p$ ranging from $-2l-3$ for detectors located near the black hole horizon up to $-l-2$ measured by the far located detectors including infinity.

\subsection{Solution for Boson Stars}

In order to compare the behavior of the scalar field on a black hole space-time with the behavior on a boson star space-time we choose boson star configurations that are stable, that is, potentially black hole mimickers \cite{GuzmanBecerril2009}. We also choose two different values of the self-interaction parameter $\Lambda=0,20$, so that we sample a representative sector of the parameter space of boson star configurations. For this we choose the four configurations described in Table \ref{table:configs}.

\begin{table}[htb]
%\centering
\begin{tabular}{|c|c|c|c|c|}
\hline
Label & $\Lambda$ & $\phi_0(0)$ & ADM Mass \\\hline
A & 0   & 0.25    &    0.632  \\
B & 0   & 0.1    &    0.533    \\
C & 20 & 0.1    &  1.39     \\
D & 20 & 0.05  &     0.8365    \\\hline
\end{tabular}
\caption{Configurations selected to compare the evolution of the test scalar field on a boson star space-time. Configuration A is the most compact one, that is, it has the highest $2M_{99}/r_{99}$ ratio, where $r_{99}$ is the radius of the 2-sphere containing 99\% of the whole mass of the configuration (see \cite{Guzman2006}); in the same sense, configuration B is a very diluted configuration. The effect of self-interaction is that more mass can be maintained by gravity at the price of a more expanded -- less compact -- configuration \cite{Guzman2006}, here configuration C is next to the critical configuration and more compact than configuration D (see Fig. 1).}
\label{table:configs}
\end{table}

\begin{figure*}[htp]
\includegraphics[width=7.5cm]{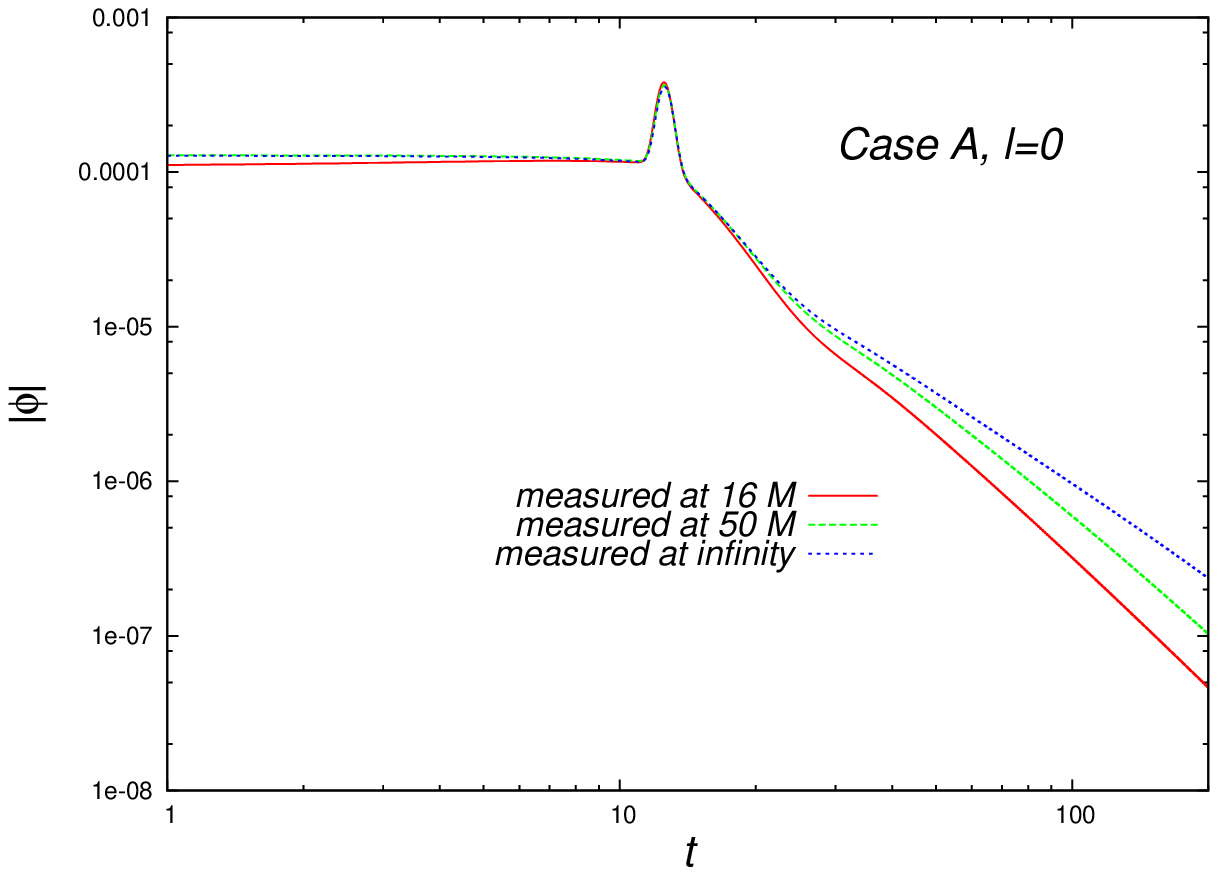}
\includegraphics[width=7.5cm]{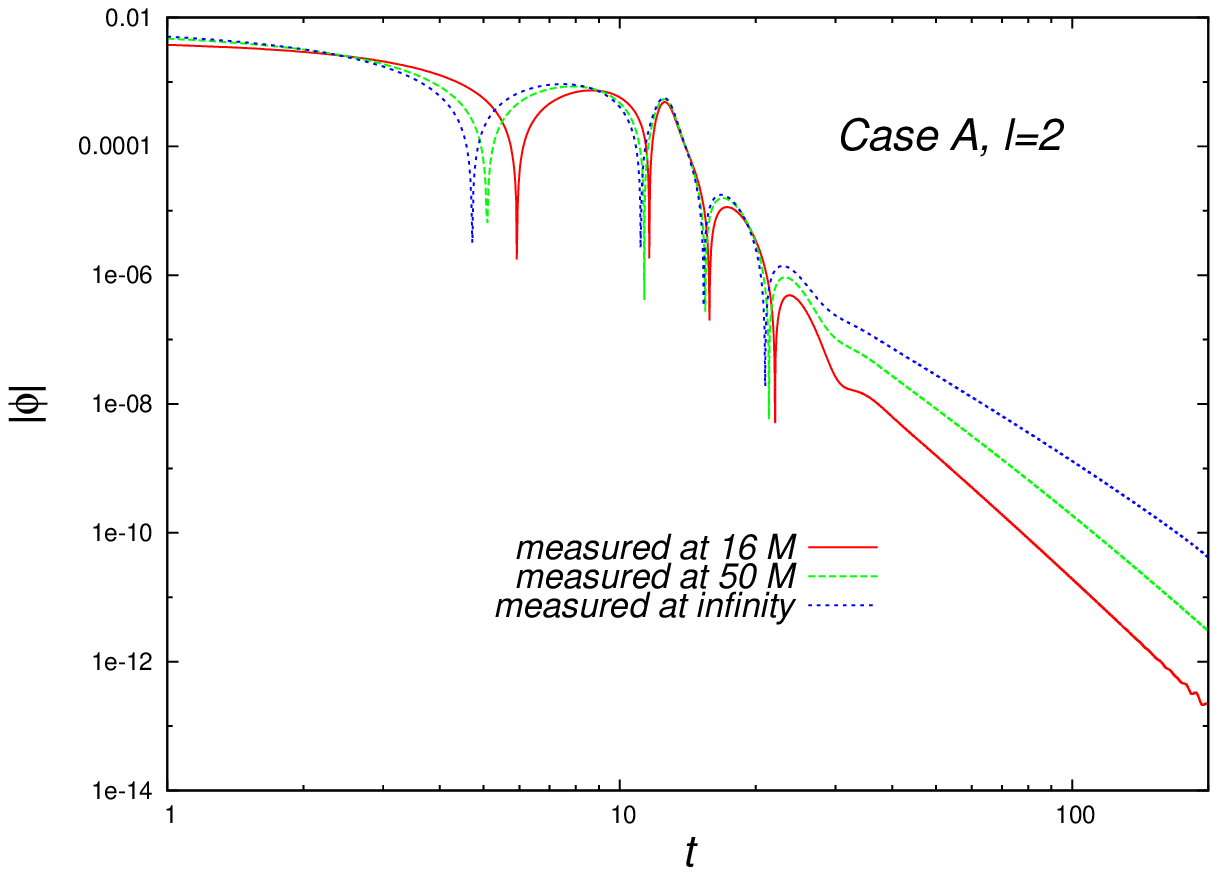}
\includegraphics[width=7.5cm]{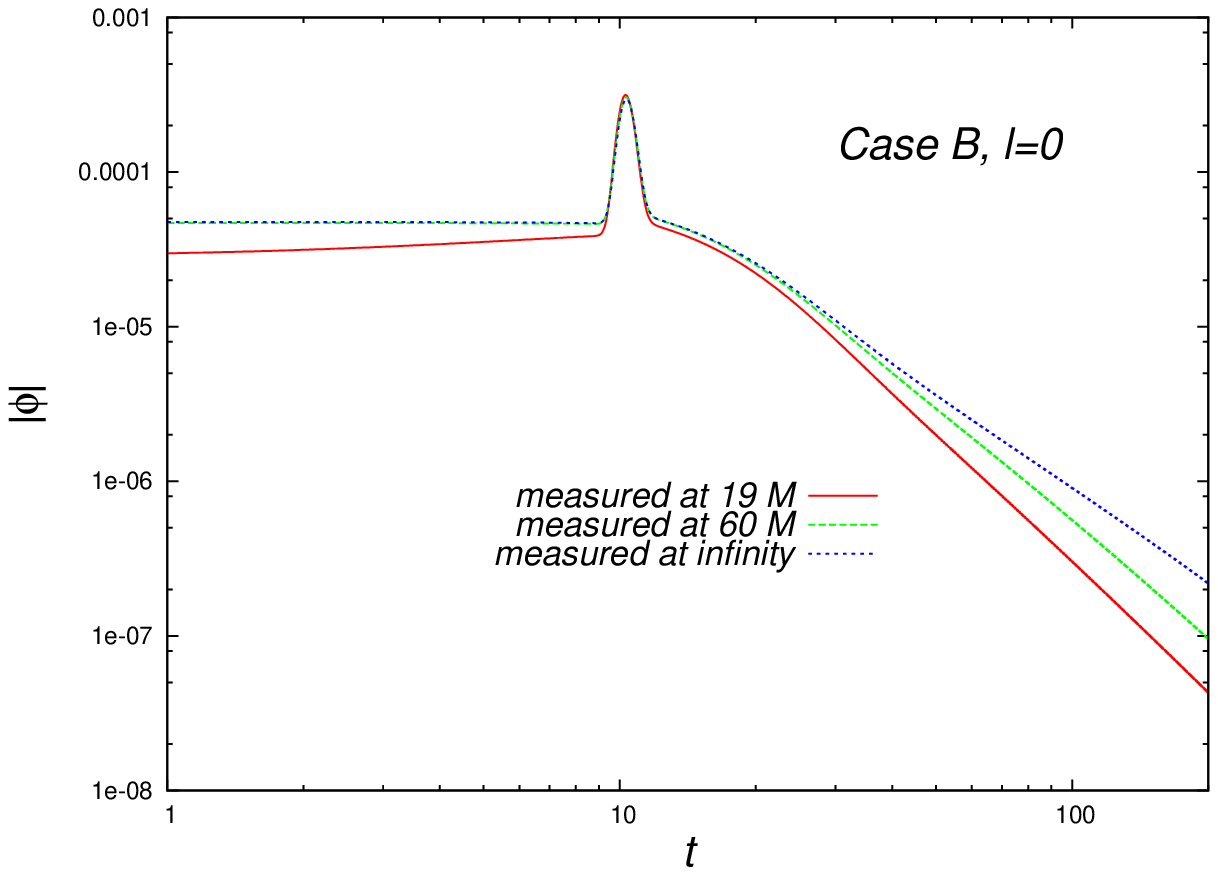}
\includegraphics[width=7.5cm]{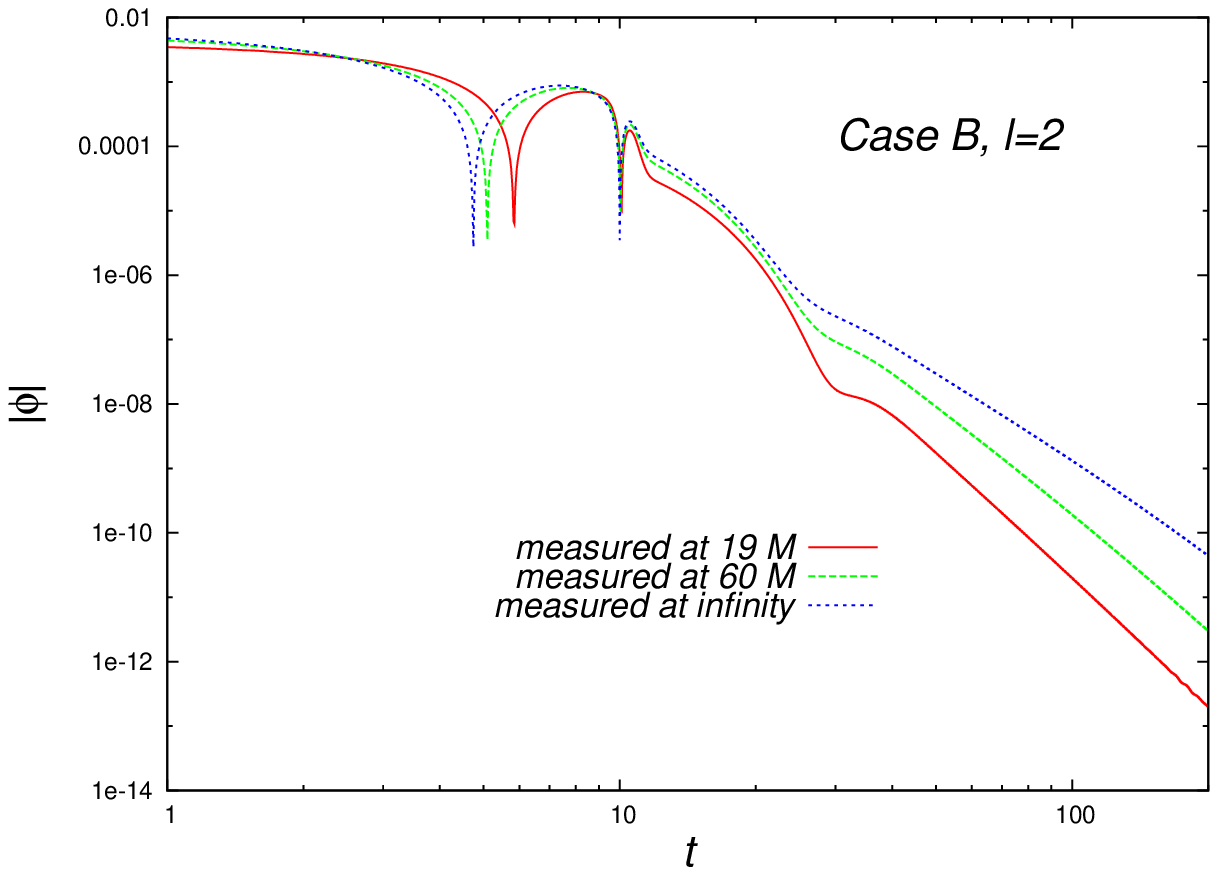}
\includegraphics[width=7.5cm]{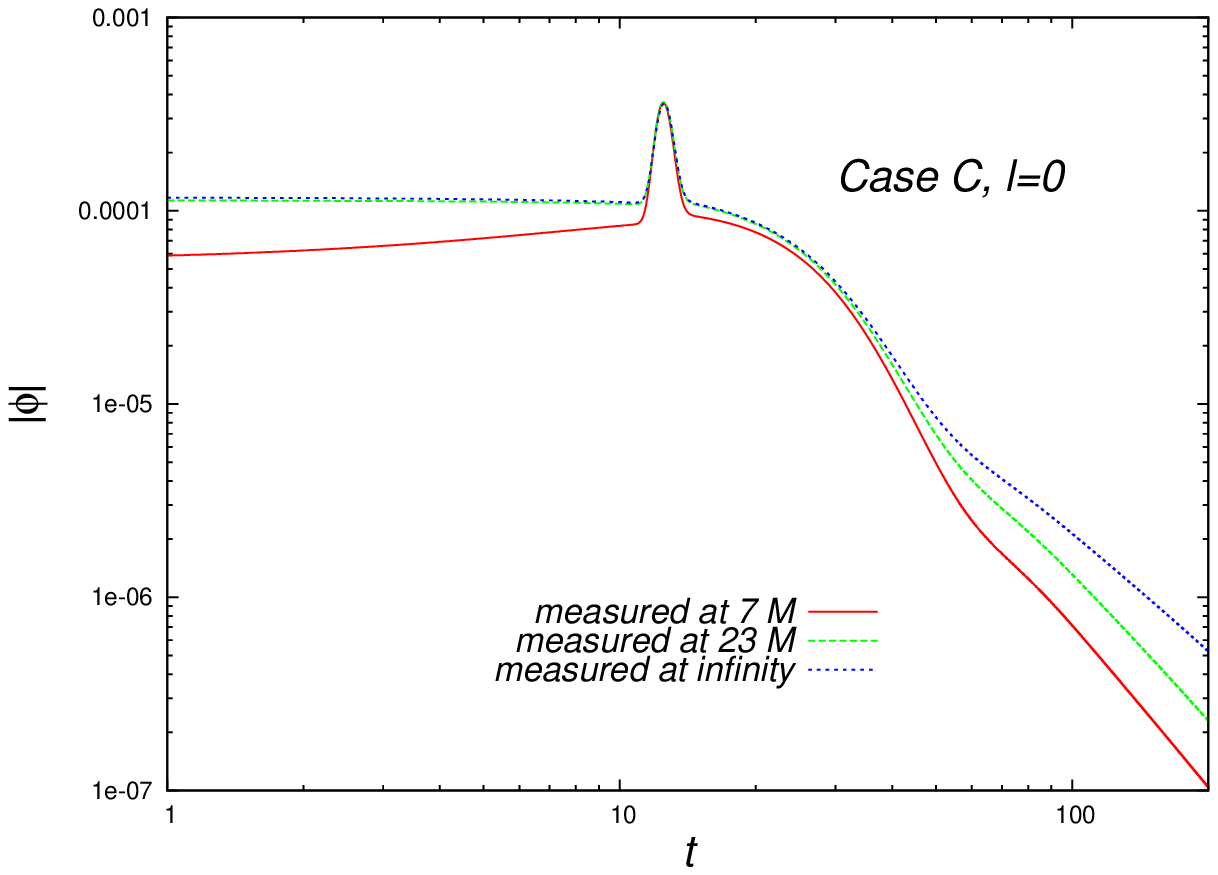}
\includegraphics[width=7.5cm]{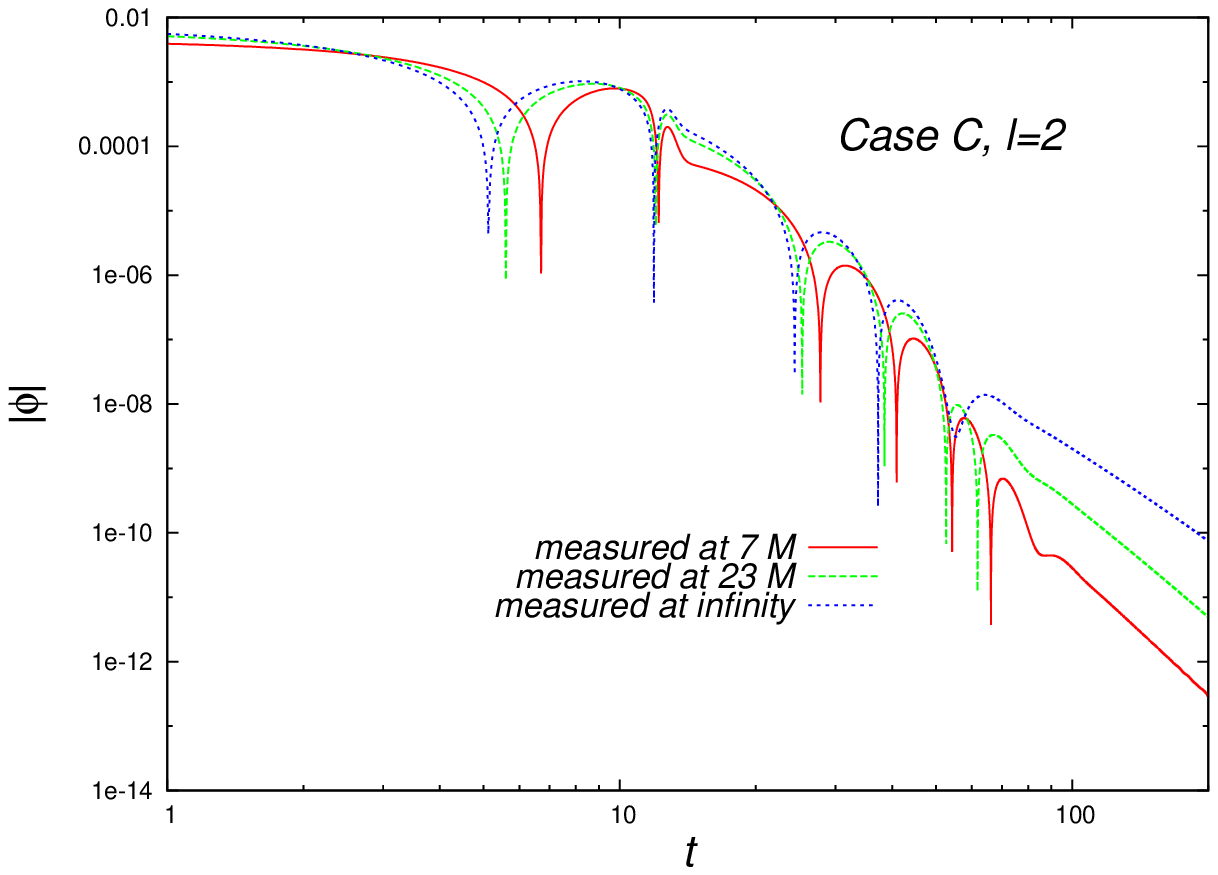}
\includegraphics[width=7.5cm]{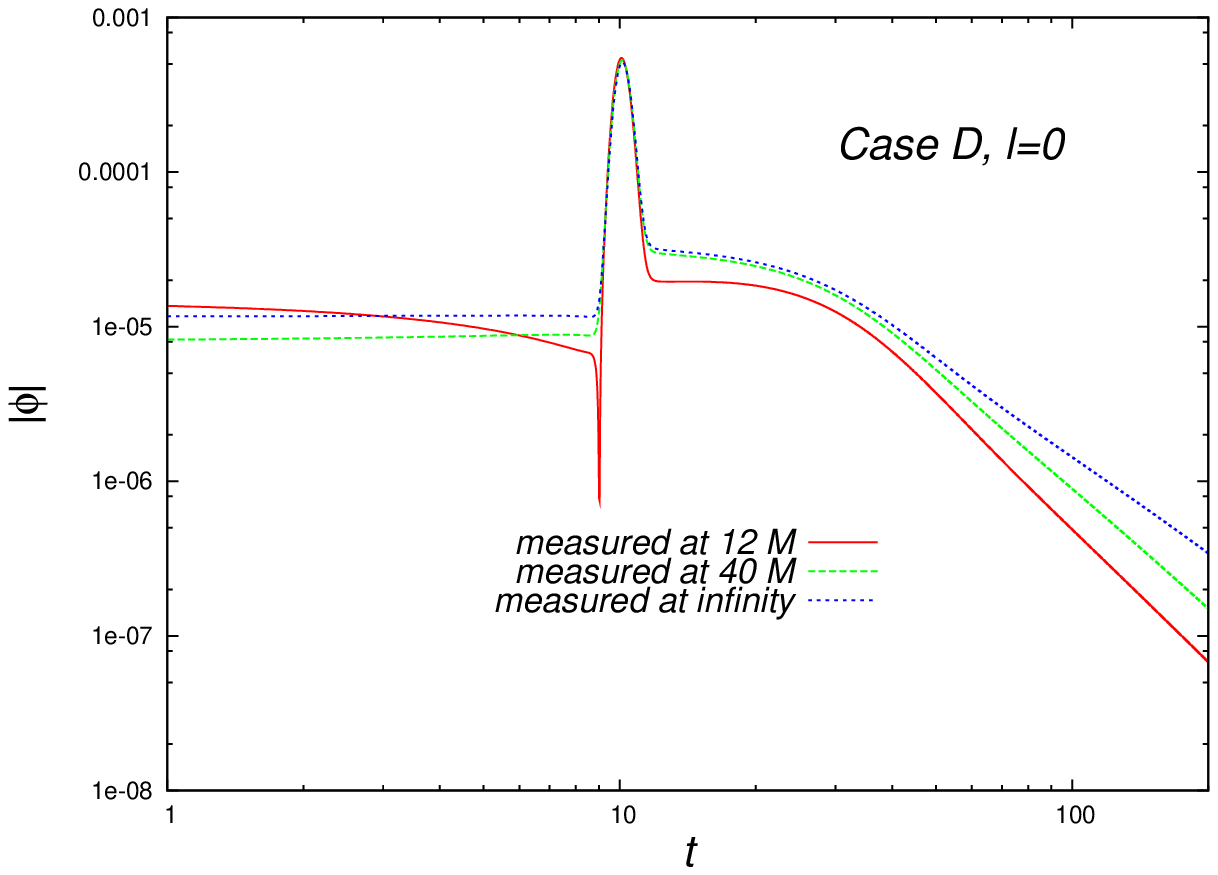}
\includegraphics[width=7.5cm]{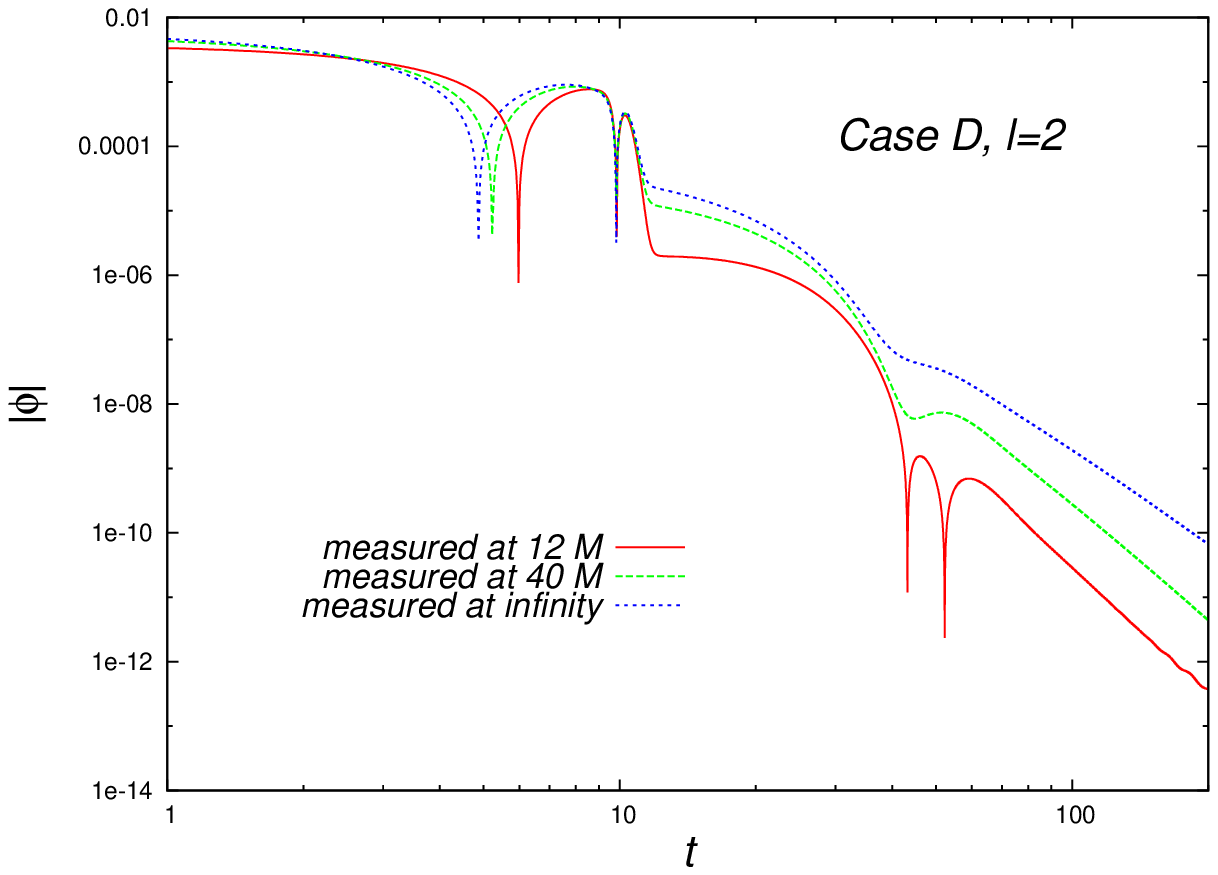}
\caption{\label{fig:exponents_bs} (First row) We show the tail decay exponents for two boson star configurations with $\Lambda=0$. In the first plot we show the scalar field amplitude corresponding to the case of central scalar field value $\phi_0 (0)=0.25$ measured at various distances from the center of the star; this boson star configuration is next to the most compact configuration which is the critical solution corresponding to $\phi_0(0)=0.254$ \cite{Guzman2006};  for $l=2$ we show a window where oscillations similar to those of quasinormal modes are observed, whereas for $l=0$ the field simple decays. (Second row) We show the same quantities for the case of a boson star with $\phi_0 (0)=0.1$ which is less compact that the other configuration. (Third and fourth rows) Results corresponding configurations with $\Lambda=20$ and $\phi_0 (0)=0.1$ nearly the most compact for this value of self-interaction and $\phi_0 (0)=0.05$ less compact.} 
\end{figure*}

We measure the amplitude of the scalar field at various fixed distances. The absolute value of the scalar field measured at various distances is shown in Fig. \ref{fig:exponents_bs} for the four configurations described in the table. 
It can be noticed in the figure that for the $l=0$ mode, the scalar field simply decays and does not show oscillations of any kind, pretty much as also found for black holes. However, for the mode $l=2$
the pair of compact configurations (cases A and C) show a set of oscillations, whereas their diluted counterparts (cases B and D) do not show a clear train of oscillations and the amplitude of the scalar field simply drops. This is an important fingerprint, and a property to cross-check the boson star parameters of boson mass and self-interaction (or compactness) used to construct black hole mimickers \cite{Guzman2006,GuzmanBecerril2009}.

\begin{figure*}[htp]
\includegraphics[width=8cm]{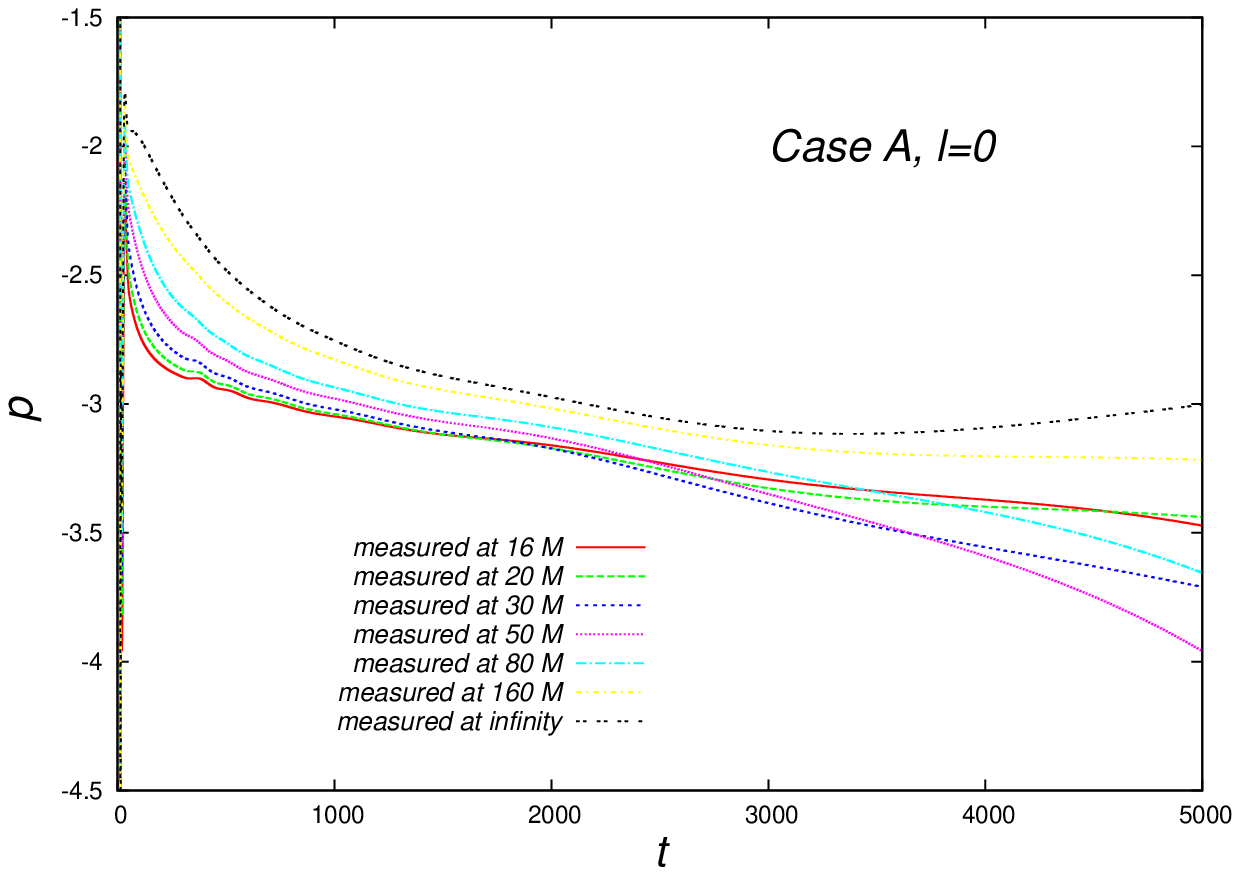}
\includegraphics[width=8cm]{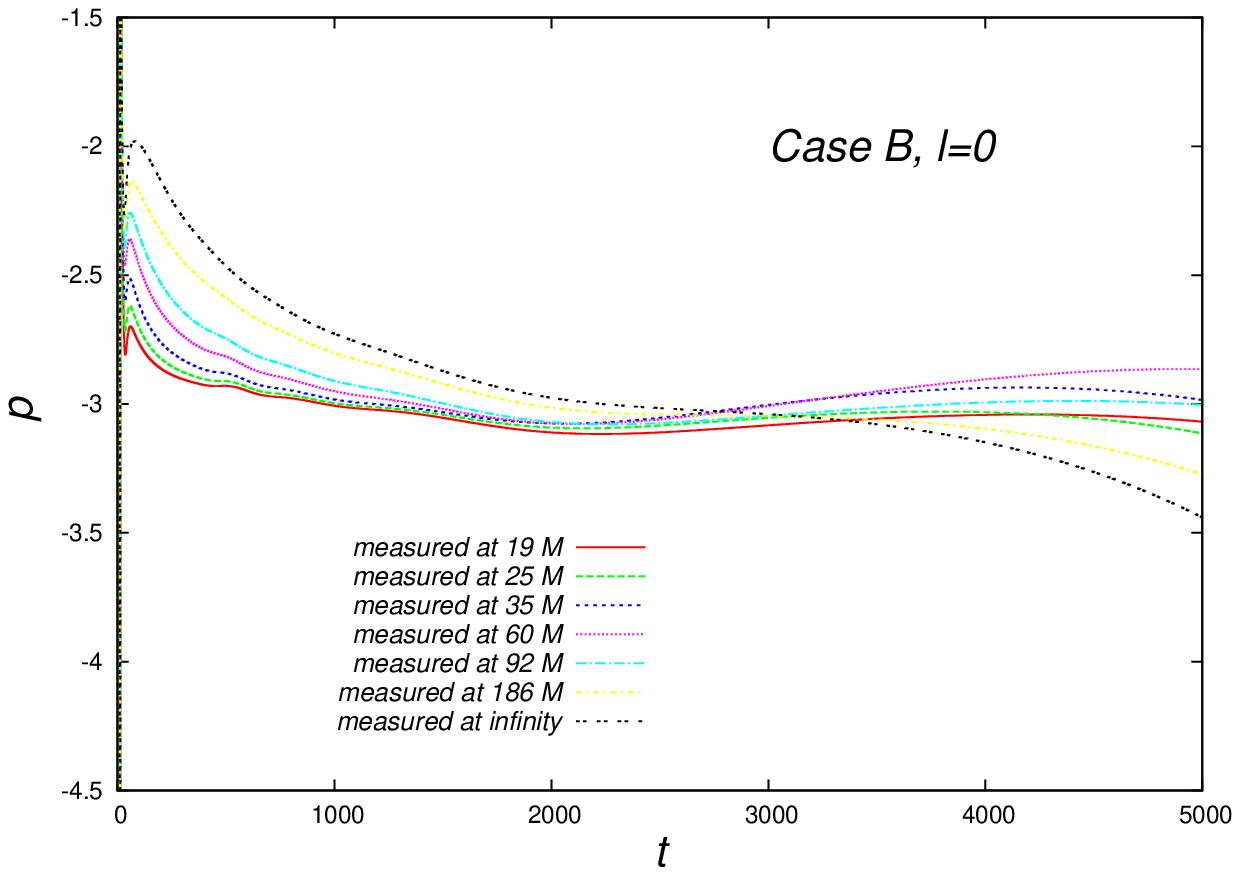}
\includegraphics[width=8cm]{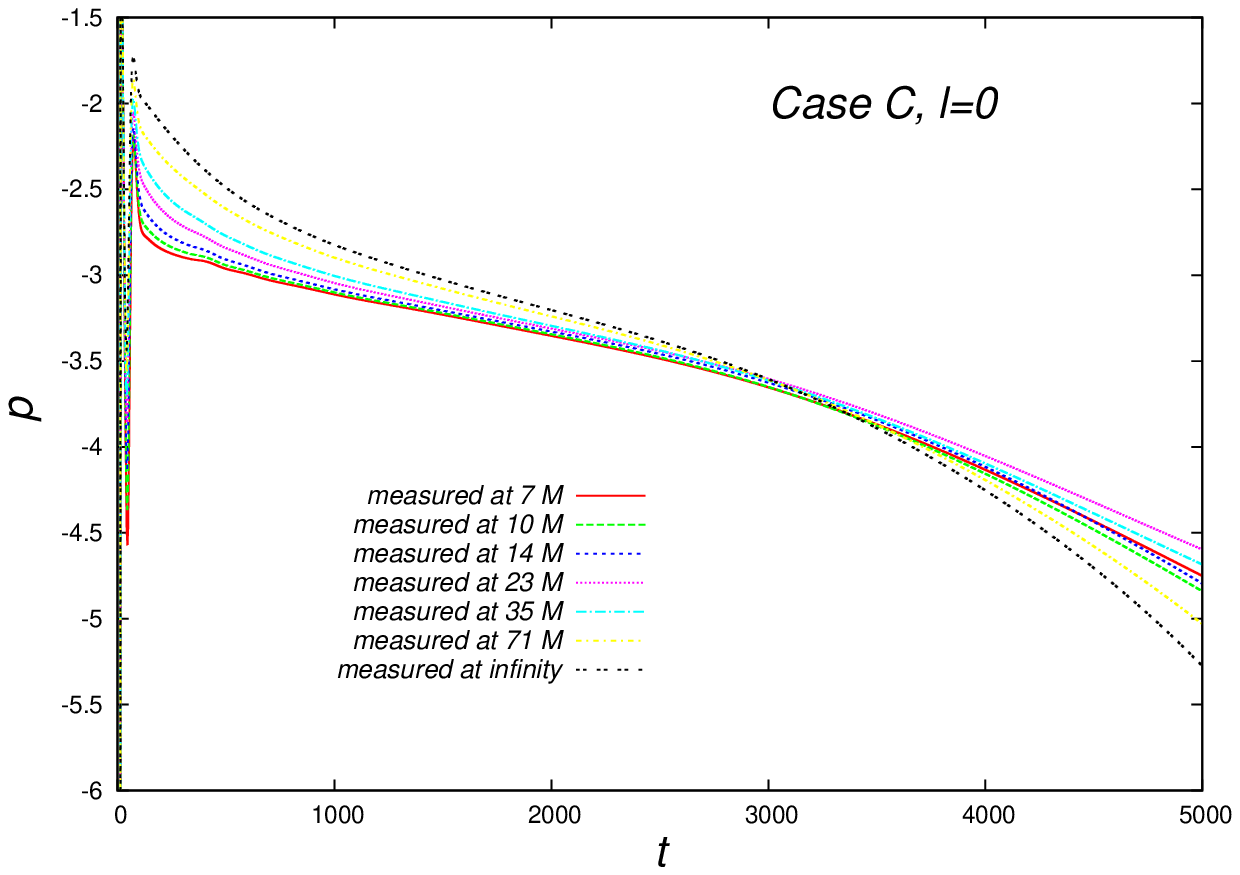}
\includegraphics[width=8cm]{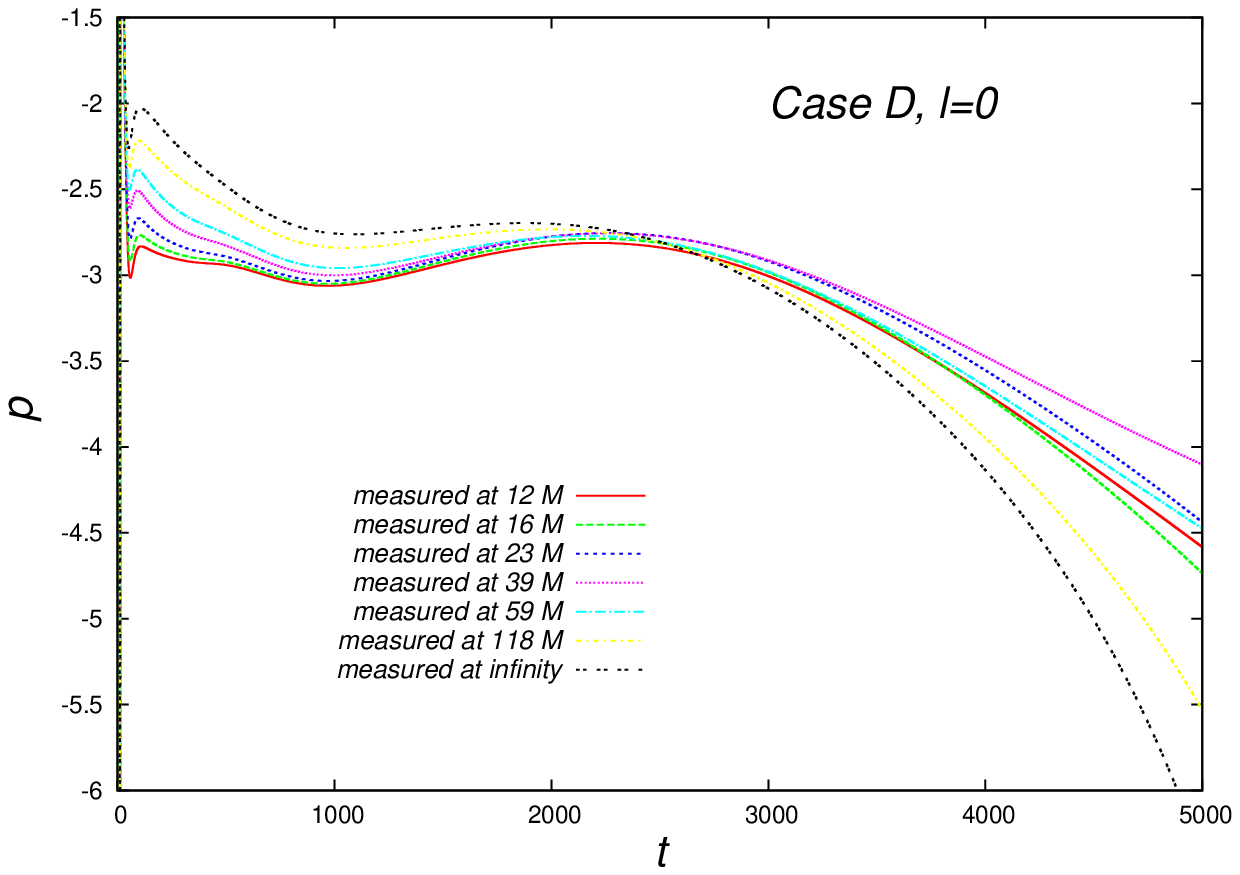}
\caption{\label{fig:bs_tails} We show the polynomial exponent $p$ measured by various observers in time for $l=0$. No signs of stabilization of $p$ were found.} 
\end{figure*}

Aside from the oscillations, we track the tail-type of behavior following the phase of oscillation of the scalar field. What  has been found for black holes, is -- as shown in Fig. \ref{fig:test1} -- that the scalar field amplitude shows a polynomial tail decay of the type $\sim t^p$ for large values of $t$, where $p$ approaches a constant negative value, and also observers located at various positions measure different tail decay exponents. We explore the same possibility for boson star space-times. Our results for $l=0$ are presented in Fig. \ref{fig:bs_tails}. For this we calculate the exponent $p$ using the fact that assuming $\phi(x,t) \sim K t^p$ with $K$ a constant for each position where the amplitude of the field is measured, the exponent can be expressed by $p = \frac{d \log |\phi|}{d \log (t)}$, and this exponent may depend on position and time as shown for the black hole case in Fig \ref{fig:test1}. We found in all the BS cases that the exponent $p$ does not stabilize around a constant value, instead, the exponent remains decreasing up to amplitudes of the wave function near to round-off error values. For $l=2$ we found similar results, although the amplitude of the scalar field approaches  round-off error amplitudes very quickly.

Finally, we have verified the convergence of our results and found that it ranges from fourth to sixth order of convergence when resolution is increased, as expected from the accuracy of our numerical methods.

% ----->     Conclusions     SECTION V     <-----

\section{Conclusions}

We have solved the massless test scalar field equation on space-times of  Schwarzschild black holes and boson star configurations. The motivation has been the quest of restrictions on the boson star parameters acting as black hole mimickers. We have found that this test field has a very different behavior when evolving on a Schwarzschild black hole space-time and when it evolves on a boson star background.

We found oscillations of the test field and signs similar to a quasinormal mode ringing for compact configurations in the $l=2$ case, whereas for diluted solutions such effect was not  as clear. Since we are solving for the massless scalar field and not for the perturbation master equation of boson stars -- which has to include the coupling of the matter the boson star is made of -- we cannot claim that these are the quasinormal modes of boson stars. However we have an indication that not all the stable BS configurations show clear oscillations for all the potential configurations that may act as black hole mimickers.

Another important sign we have found within the precision of our algorithms is that for boson stars, even though they correspond to a space-time with curvature, no polynomial tail decay was found -- at least we did not track a stable polynomial exponent. We consider this to be an very important fingerprint of gravitational radiation detected once tails can be measured.

% ----->     ACKNOWLEDGMENTS     <-----

\section*{Acknowledgments}

The authors thank A. Zenginoglu for important discussions on the scri-fixing conformal compactification.
This research is partly supported by Grant Nos.
CIC-UMSNH-4.9 and 
CONACyT 106466.
The runs were carried out in the IFM Cluster.

% -------------------------------------------------------
% -----     REFERENCES     ----------

\end{document}